\documentclass[11pt,twoside]{article}
\usepackage{newpasp}
\usepackage{epsf}
\begin{document}
\title{Large-scale Gas Dynamics in the Milky Way}
\author{Peter Englmaier \& Ortwin Gerhard}
\affil{Astron. Institut Uni Basel, Venusstr. 7, 4102 Binningen, Switzerland.}
\author{Nicolai Bissantz}
\affil{Institut f\"ur Mathematische Stochastik, Universit\"at
G\"ottingen, Lotzestr.~13, 37089~G\"ottingen, Germany.}
\begin{abstract} 
We present new gas flow models for the Milky Way inside the solar
circle. Using a mass model derived from the COBE/DIRBE maps and clump
giant star counts, and using a parametric model for the spiral arm
pattern in the disk, we calculate the gas flow and compare with $^{12}$CO
observations.  We find that models with 4 spiral arms fit the
observations better then 2-armed models. We also find that models
with separate pattern speeds for the bar and spiral arms can explain
the gas flow in the bar corotation region better than single-pattern
speed models.
\end{abstract}
\section{Introduction}
The Milky Way (MW) is the best studied galaxy and is particularly
interesting for studies of galactic structure such as bars and spiral
arms and dynamics. While it has been notoriously difficult to assess
the MW's large scale structure and dynamics because of our position
within the galactic plane, a rich amount of small scale details will
allow to test theories of galaxy dynamics and evolution once this
problem is solved. One longstanding problem is the origin and dynamics
of spiral arms. In particular for the MW: what is the shape and
amplitude of the spiral arms?  Do they rotate with a fixed pattern
speed? Is the bar the driving force behind the spiral pattern, or do
bar and spiral arms have multiple pattern speeds?

\section{Models for the mass distribution}

Models for the mass distribution of the Milky Way have been
constructed based on parametric fits for disk and bulge to the
observed NIR luminosity distribution with a constant mass-to-light
ratio for each component, and an analytical description of the dark
matter halo. Perhaps the most successful axisymmetric model has been
given by Kent~(1992), who used the $2.4\,\micron$ (K-Band) IRT map to
fit parametric models for disk and bulge and compared to the rotation
curve derived from atomic hydrogen 21-cm maps.

This basic approach was also followed by recent studies using the
higher quality COBE/DIRBE near-IR maps which open new possibilities
for studies of non-axisymmetric structure in the bulge and inner
disk. The peanut shape of isophotes in the bulge has been attributed
to a bar or triaxial bulge (Dwek et al. 1995; Binney, Gerhard, \&
Spergel 1997; Freudenreich 1998), which has been claimed to exist
since the early galactic 21~cm maps (see in Gerhard 1996).

Unlike distant galaxies, our view of the inner Galaxy is subject to
perspective effects, which can be used to infer its true 3-dimensional
light distribution, and has inspired development of non-parametric
deprojection methods. One key ingredient for these methods is the presence
of a bar, which is assumed to be nearly 8-fold symmetric. 

The quality of such models is limited by 
angular resolution and depends on the quality of the underlying dust
correction model, such as those by Spergel, Malhotra, \& Blitz (1996), Drimmel
\& Spergel (2001). In addition, noise and non-uniqueness limit the
predictive power of the deprojected model (Binney et al. 1997; Zhao
2000).

\begin{figure}
\plotone{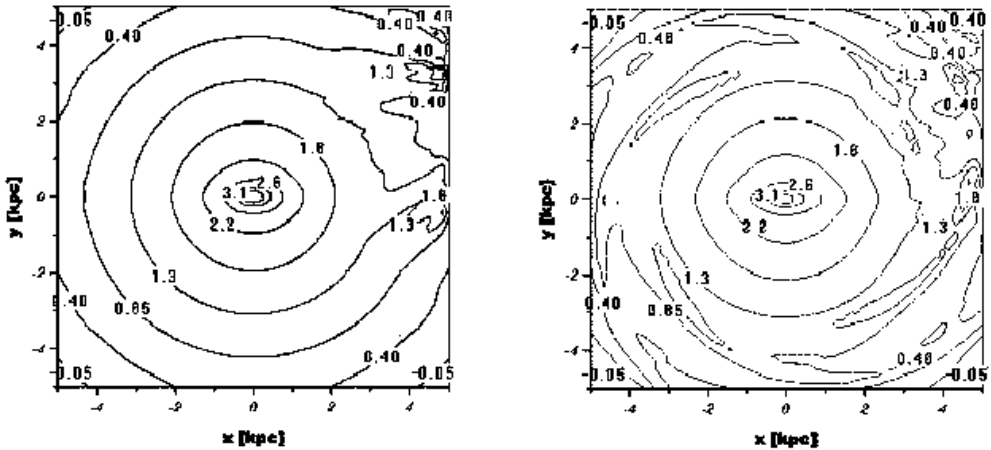}
\caption{%
Recovered face-on surface density of the MW model from 
Bissantz \& Gerhard (2002). Left: best model neglecting spiral arm structure. 
Right: best model biased toward a parametric spiral arm model close to the 
$z=0$ plane.
}
\end{figure}
The recovered 3D-structure of one such model is shown in Fig.~1,
derived by Bissantz \& Gerhard (2002). Their method attempts
to iteratively estimate the luminosity distribution in the
inner Galaxy from the COBE/DIRBE $L$-band data, while maximizing a 
penalized likelihood function.
This method is non-parametric in the sense that it can adjust the
luminosity of approx.\ 150 000 grid points, which
cover the inner 5 kpc of the MW, such that the overall fit to the
COBE near-IR map is optimized, that is, no functional form of the
density distribution is assumed.

The likelihood function is biased by 3 penalty terms: First, the
luminosity density function should be smooth. This is to suppress the
amplification of noise in the data. Second, the Galaxy should be
approximately 8-fold symmetric. This symmetry is fulfilled by a bar-like
distribution, however, spiral arms are 4-fold symmetric.
Therefore, a third bias is introduced, namely the model should be
close to a given spiral arm model in the galactic plane. 

The method attempts to balance between all constraints, and
there are statistical arguments and tests which demonstrate that the
result is dominated by data and not by one of the penalizing criteria 
(Bissantz, Munk, \& Scholz 2003).

\begin{figure}
\plotone{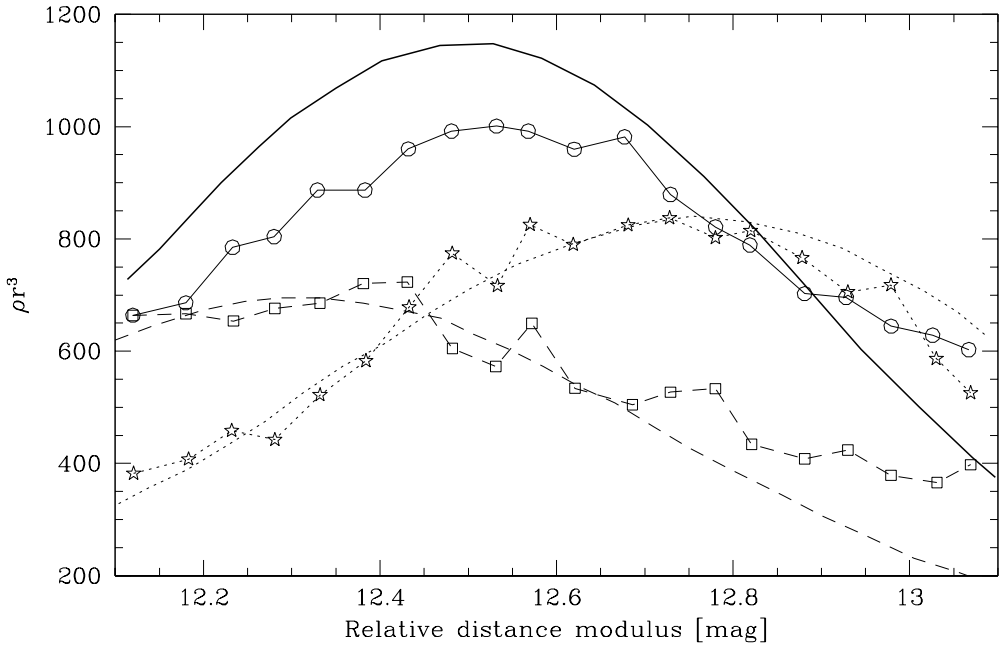}
\caption{%
Model results from Bissantz \& Gerhard (2002) compared to line-of-sight 
distributions of clump giants observed by Stanek et al.~(1994, 1997). 
The bar in the model with spiral arms explains the observed asymmetry
nicely. Solid, dotted, and dashed curves are fits to the data points 
for Baade's window, $l=-4.9\deg$, and $l=5.5\deg$. 
The smooth curves without symbols are model results.
}
\end{figure}
For a given position of the sun with respect to the bar, a best model
is found. The non-parametric model is currently limited to the central
5~kpc of the Galaxy. Further out, the model is continued by a
parametric model, while at the very center, the density profile is
modified to include a cusp.

When comparing the COBE/DIRBE light distribution to the projected
model, with or without spiral arms, it fits almost perfectly. Hence,
the bar in the galactic center is sufficient to explain the asymmetry
in the near-IR light; no clear direct evidence for spiral arms is
found in the data.  Indirect evidence rather comes from the fact that
artificial features in the disk plane are greatly reduced when spiral
arms are taken into account (see Fig.~1). More recent studies by
Drimmel \& Spergel (2001) find weak evidence for a spiral arm pattern in the
DIRBE maps when the bar region is excluded.

To improve the disentanglement of bar and spiral arm
contributions, additional data is needed. One very useful data set
comes from the line-of-sight distribution of red clump giants. Stanek
et al.~(1994, 1997) observed three fields toward the bulge and found
that on the left side of the bulge the maximum of the red clump giant
density is nearer than on the right side.
Models with spiral arms yield a more elongated bar,
which can explain the asymmetry between positive and negative
\begin{figure}[!t]
\plotone{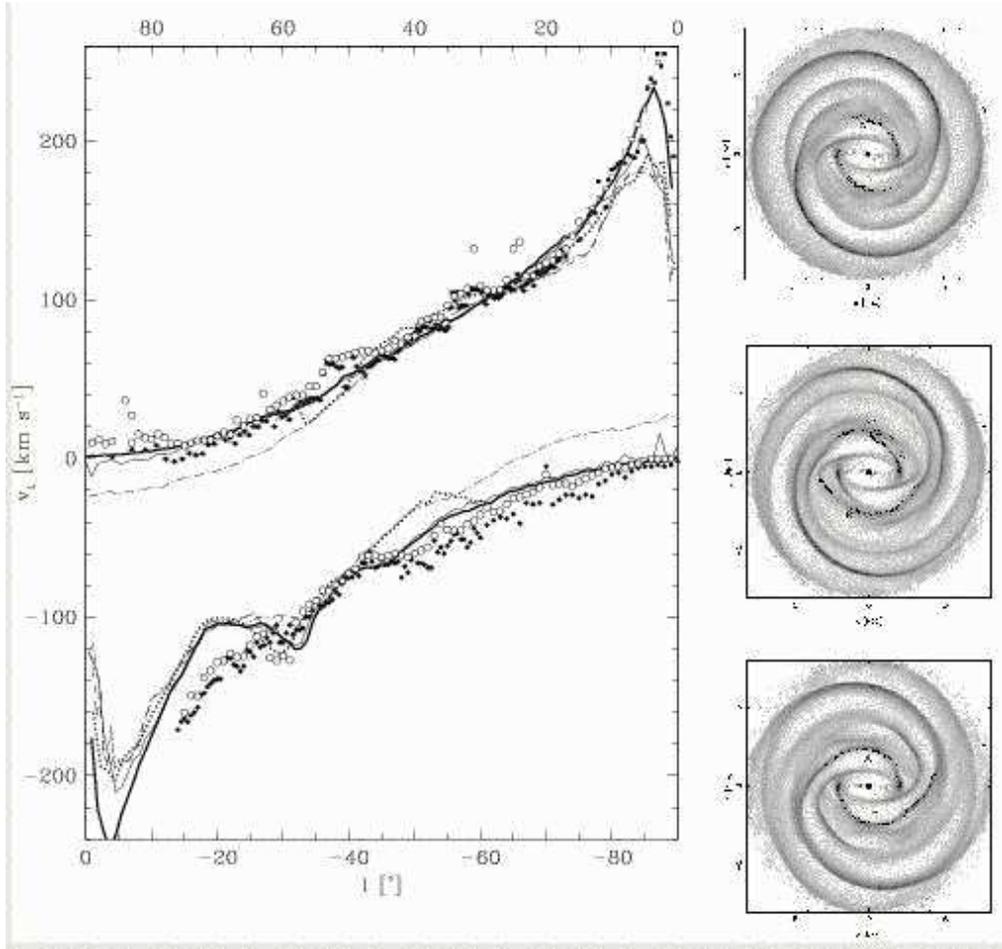}
\caption{%
Left: Model terminal rotation curve (thick solid curve) compared to 
observations (symbols); for more details see Englmaier \& Gerhard (1999). 
Right, from top to bottom: 
MW gas flow model from Bissantz et al.~(2003) with spiral arm pattern
speeds 20, 40, and 60~km/s/kpc. The position of the Sun is indicated by a
circle in the upper right.
}
\end{figure}
longitudes in the red clump distribution quite nicely, while models
without spiral arms fail to do so (see Fig.~2). Note, that the spiral
arms do not contribute significantly themselves to the red clump distribution
asymmetry, they merely modify the shape of the deprojected bar.

Another successful model of the mass distribution was found by
Fux (1997). His fully self-consistent model was made using bar
formation unstable n-body models. By selection from a sample of
models, he was able to find a model and time step which is very similar
to the observed inner galaxy (residual 0.4\%). The precision of this 
method is limited
by the number of models one can run and also by the parametric
description of the initial axisymmetric n-body model. Nevertheless,
further progress in this direction will help to understand the origin
of the galactic bar and its fate.

\section{Gas  dynamics}

Observations of the gas dynamics using Doppler-shifted mm-lines offer
a direct way to test models. The Dynamics and distribution of
molecular gas in the MW are mostly governed by its dissipative nature
and the gravity of the mass distribution. The cold gas orbits around
the center of the galaxy, but streamlines are not allowed to cross and
shocks modify the gas flow to avoid such crossings. Hence, gas orbits
probe different aspects of the underlying potential from stellar
orbits.  A good mass distribution model for the MW should allow us to
reproduce the observed gas distribution and velocities.

Englmaier \& Gerhard (1999) modeled the MW gas flow using SPH, but
without a spiral arm pattern in the deprojected mass model. From the
model they computed terminal velocity curves and compared them to the
observed data.  As can be seen in Fig.~3, the fit is quite good and
deviations from the observed velocities at most longitudes are of the
order of the scatter in the observed data. The terminal curve reflects
both the axisymmetric part of the rotation and the streaming
velocities induced by in the bar. At radii near the solar radius the
dark matter halo starts to contribute significantly to the observed
terminal velocities. Hence, it can already be concluded, that the
deprojected luminosity distribution together with the assumption of
constant mass-to-light ratio is quite capable of reproducing the
correct mass distribution in the inner galaxy.

Bissantz, Englmaier, \& Gerhard (2003) extended the model by including
spiral arms in the mass model. They confirmed the basic parameters for
the galactic bar found by Englmaier \& Gerhard (1999), i.e.\ pattern
speed $60\,$km/s/kpc, corotation radius $3.4\,$kpc, orientation angle
$20\deg$, but also allowed for 3 different values for the fixed
pattern speed of the spiral arms, namely 20, 40, and 60 km/s/kpc.  In
the last model, the spiral pattern corotates with the bar.

\begin{figure}[!t]
\plotone{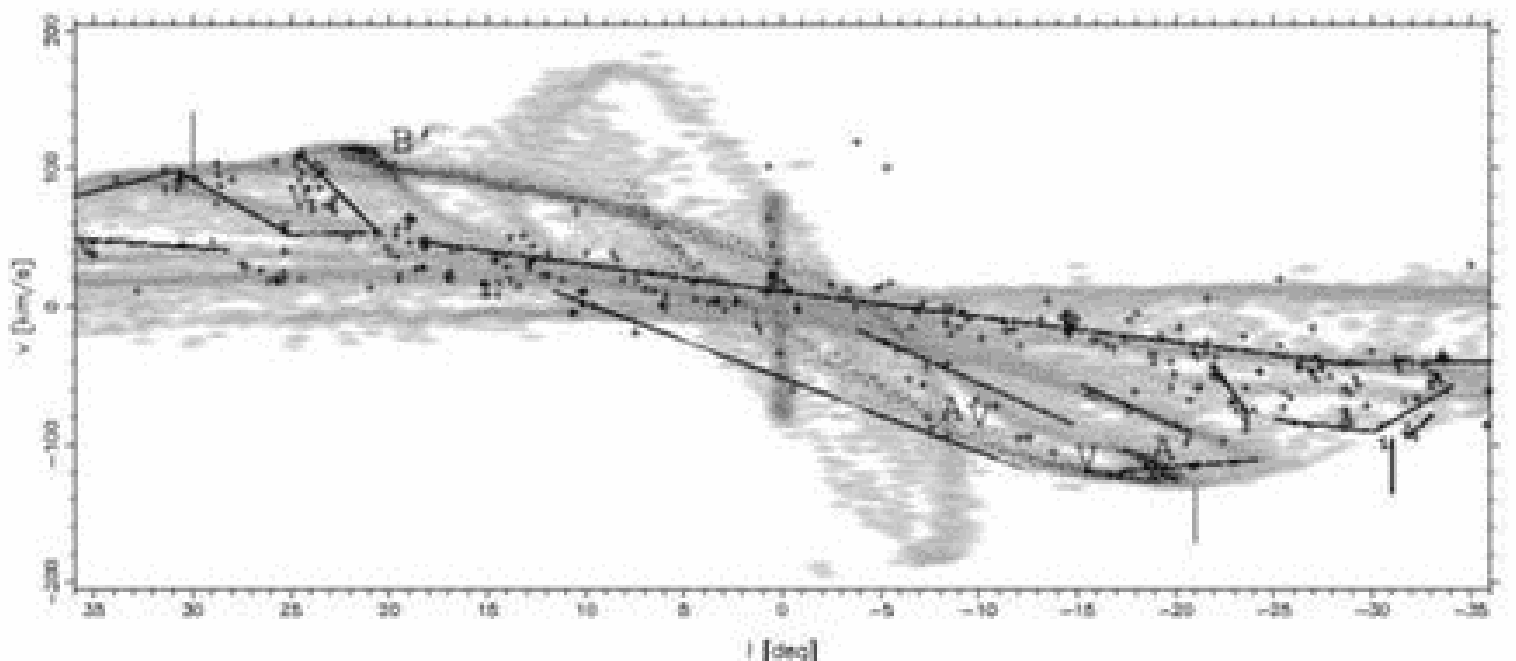}

\plotone{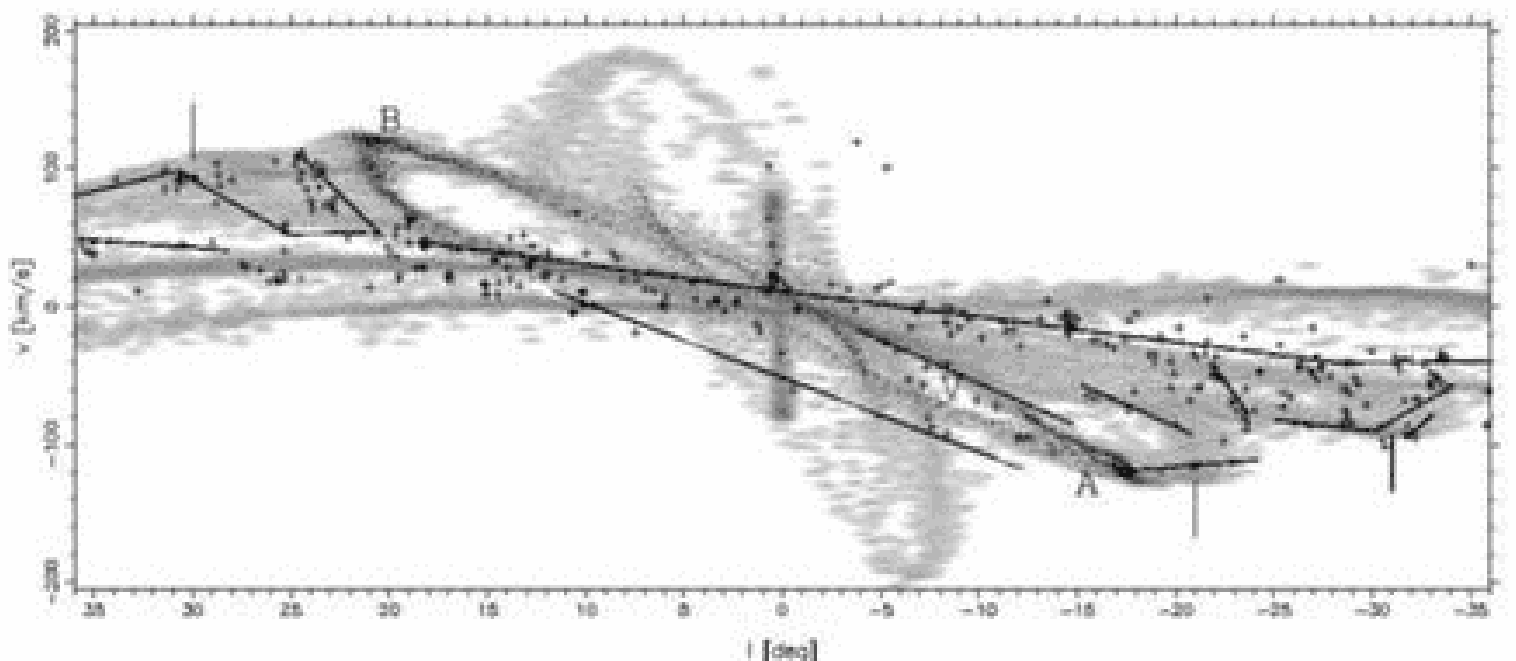}

\plotone{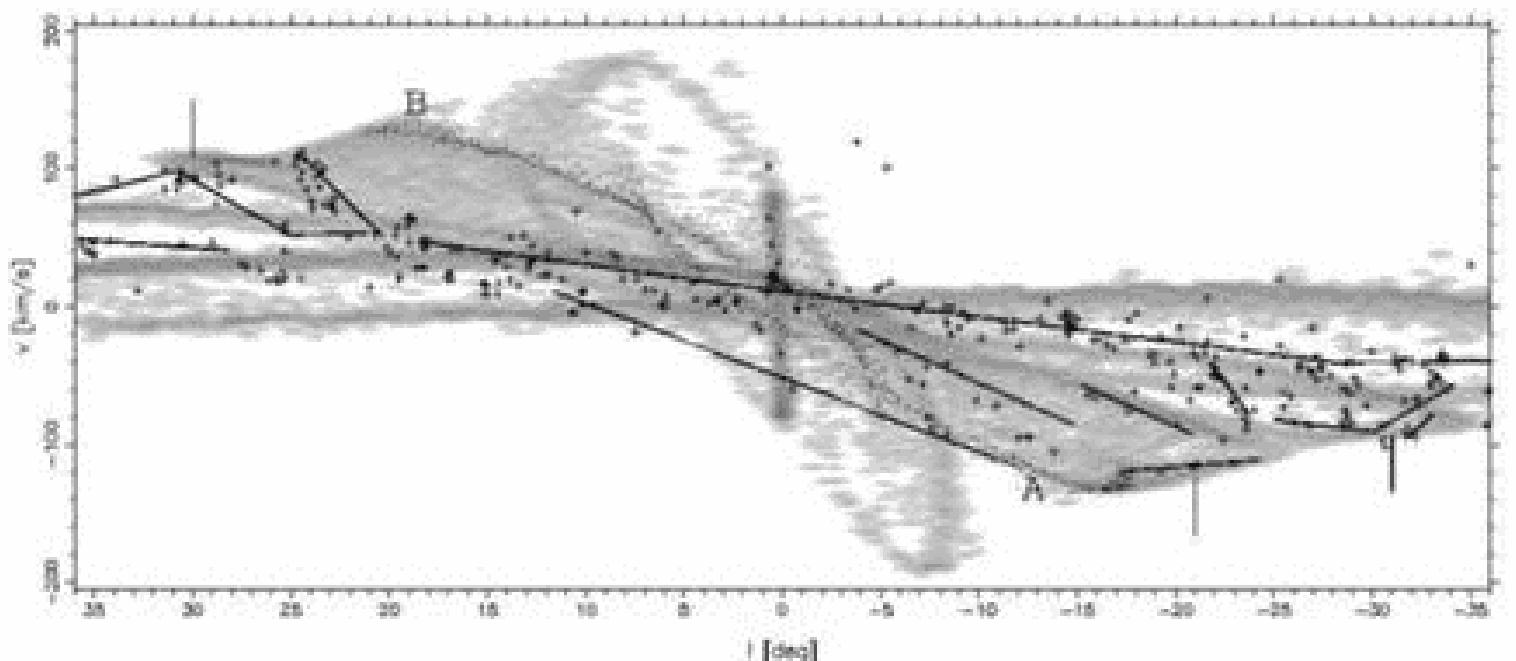}
\caption{%
Lv-diagrams for the models shown on the right in Fig.~3. The
spiral arm pattern speed is 20, 40, and 60~km/s/kpc (from top to bottom). 
The voids mentioned in the text are indicated by the letter `V'.
}
\end{figure}
In these gas models, at almost any given time, the outer spiral arms
smoothly connect to the inner spiral arms. See Sellwood \& Sparke
(1988) for a similar finding. For any model with a given pair of 
pattern speeds, a best model time can be found by comparison with 
the observed gas dynamics.

One can see in these 3 cases, that the spiral pattern differs mostly
in the transition region around the corotation radius of the bar,
where both the bar and the spiral  arms impose perturbations with
different pattern speeds. The solution, of course, is time-dependent
but nearly periodic. In Fig.~3 time steps which are closest to the observed
velocity data are shown.  Note, that the spiral arms disappear in the
corotation region in the model with a single pattern speed while they
pass smoothly through the corotation region of the bar in other models.

Bissantz et al.~(2003) calculated longitude-velocity (lv) diagrams
from the SPH models to compare with surveys of the molecular gas
($^{12}$CO), but also HII regions and giant molecular clouds.  One
difficulty with lv-diagrams is that a coherent spatial structure does
not always correspond to a clear structure in the lv-diagram and vice
versa.  In general the models reproduce the large scale structure
qualitatively and quantitatively quite well, although some features
related to an asymmetric mass and gas distribution cannot be
reproduced correctly.

In the lv-diagram region which corresponds to the corotation region of
the bar, the $^{12}$CO molecular gas distribution observed by Dame et
al.\ clearly shows spiral arm ridges and void areas with low gas
content.  In the models, Bissantz et al.~(2003) find similar voids in
this region, but only if the spiral pattern rotates slower then the
bar pattern (Fig.~4).  In the model where the spiral and bar pattern
rotate with the same speed, no voids form.  The reason is that spiral
arms cannot propagate through their own corotation region, because the
solution for density waves is exponentially decaying there (Goldreich
\& Tremaine 1978).  Hence, the spiral arms smear out and leave no
clear signature in the lv-diagram.

Although a 4--armed pattern in the MW is preferred by many papers,
some data seem to indicate a 2--armed pattern (Vall\'ee 1995).
Unfortunately, models with 2 spiral arms are also possible with the
DIRBE K-band data, while the dust seen in the 240~\micron\ data
prefers a 4-armed pattern (Drimmel \& Spergel 2001).  Bissantz et
al.~(2003) find that 2--armed models produce too few spiral arm ridges
in the lv-diagram and they poorly match the observed spiral arm
ridges. In other words, if the COBE/DIRBE data is combined with
molecular gas data dynamics, a 4--armed pattern is preferred.

\end{document}